\documentclass{article}

\usepackage[preprint, nonatbib]{neurips_2025}

\usepackage[utf8]{inputenc}
\usepackage[english]{babel}
\usepackage{hyperref}

\usepackage{amsmath,amssymb,amsthm, bbm}

\usepackage{microtype} 

\usepackage{blindtext}

\usepackage{bm} 
\usepackage{graphicx}
\usepackage{url}
\usepackage{booktabs} 
\usepackage{enumitem} 
\usepackage{hyperref} 
\usepackage{xcolor} 
\usepackage{stmaryrd}
\usepackage{bm}
\usepackage{multicol, multirow}

\usepackage{listings}

\usepackage{xparse}

\usepackage{booktabs}
\usepackage{caption}        
\NewDocumentCommand{\codeword}{v}{%
\texttt{\textcolor{blue}{#1}}%
}

\usepackage{ifthen}
\newboolean{mybool}
\setboolean{mybool}{true} 

\newboolean{public}
\setboolean{public}{true}
\usepackage{enumitem}
\usepackage{wrapfig}
\usepackage{subcaption} 

\usepackage{todonotes}

\newcommand{\Lim}[1]{\raisebox{0.5ex}{\scalebox{0.8}{$\displaystyle \lim_{#1}\;$}}}

\mathchardef\mhyphen="2D

\newcommand{\msf}[1]{\mathsf{#1}}

\newcommand{\makevect}[1]{{\ensuremath{\boldsymbol{#1}}}}

\renewcommand{\vec}[1]{\makevect{#1}}

\newcommand{\F}{\mathbb{F}}
\newcommand{\Z}{\mathbb{Z}}

\newcommand{\sh}[1]{\llbracket #1 \rrbracket}
\newcommand{\arith}[1]{\llbracket #1 \rrbracket]}

\newcommand{\ang}[1]{\langle #1 \rangle}

\newcommand{\Prot}[1]{\ensuremath{\Pi_{\mathsf{#1}}}}
\newcommand{\PLogReg}{\Prot{LogReg}}
\newcommand{\PMult}{\Prot{Mult}}
\newcommand{\PSig}{\Prot{Sigmoid}}

\newcommand{\Func}[1]{\ensuremath{\mathcal{F}_{\mathsf{#1}}}}
\newcommand{\FMPC}{\Func{MPC}}

\newcommand{\figref}[1]{Fig.~\ref{fig:#1}}
\renewcommand{\eqref}[1]{(\ref{eq:#1})}

\newcommand{\figlab}[1]{\label{fig:#1}}

\tikzstyle{block} = [rounded rectangle, draw,
    minimum width=2.5cm, text centered, minimum height=1cm, node distance=1.5cm]
\tikzstyle{arrow} = [draw, -latex]

\usepackage{tcolorbox}
\tcbuselibrary{skins,breakable}

\newtcolorbox{protobox}[2][]{%
  enhanced,
  title        = {#2},
  attach boxed title to top left={xshift=+3mm,yshift*=-3mm},
  breakable    = false,
  colback      = white, 
  colframe     = black!75,
  fonttitle    = \bfseries,
  colbacktitle = black!10!white,
  coltitle     = black,
  #1
}

\newenvironment{protofig}[3]{
  \begin{figure}[!h]
    \newcommand{\FigCaption}{#2}
    \newcommand{\FigLabel}{#3}
    \begin{protobox}{#1}
    }{%
    \end{protobox}
    \caption{\FigCaption}
    \figlab{\FigLabel}
  \end{figure}
}

\usepackage{tcolorbox}
\tcbuselibrary{skins,breakable}

\makeatletter
\newcommand{\DrawLine}{%
  \begin{tikzpicture}
  \path[use as bounding box] (0,0) -- (\linewidth,0);
  \draw[color=black!75,dashed,dash phase=2pt]
        (0-\kvtcb@leftlower-\kvtcb@boxsep,0)--
        (\linewidth+\kvtcb@rightlower+\kvtcb@boxsep,0);
  \end{tikzpicture}%
  }
\makeatother

\newtcolorbox{mybox}[2][]{%
  enhanced,
  title        = {#2},
  attach boxed title to top left={xshift=+3mm,yshift*=-3mm},
  breakable    = true,
  colback      = black!4,
  colframe     = black!75,
  fonttitle    = \bfseries,
  colbacktitle = black!10!white,
  coltitle     = black,
  #1
}

\newtcolorbox[auto counter]{functionality}[2][]{%
  enhanced,
  title        = {Functionality~\thetcbcounter: #2},
  attach boxed title to top left={xshift=+3mm,yshift*=-3mm},
  breakable    = true,
  colback      = yellow!4,
  colframe     = black!75,
  fonttitle    = \bfseries,
  fontupper    = \small,
  fontlower    = \small,
  colbacktitle = yellow!10!white,
  coltitle     = black,
  #1
}

\newtcolorbox[use counter=pro]{protocol}[2][]{%
  enhanced,
  title        = {Protocol~\thetcbcounter: #2},
  attach boxed title to top left={xshift=+3mm,yshift*=-3mm},
  breakable    = true,
  colback      = black!4,
  colframe     = black!75,
  fonttitle    = \bfseries, 
  fontupper    = \small,
  fontlower    = \small,
  colbacktitle = black!10!white,
  coltitle     = black,
  #1
}

\newtcolorbox[use counter=pro]{procedure}[2][]{%
  enhanced,
  title        = {Procedure~\thetcbcounter: #2},
  attach boxed title to top left={xshift=+3mm,yshift*=-3mm},
  breakable    = true,
  colback      = cyan!2,
  colframe     = black!75,
  fonttitle    = \bfseries, 
  fontupper    = \small,
  fontlower    = \small,
  colbacktitle = cyan!5!white,
  coltitle     = black,
  #1
}

\newtcbox{\xmybox}[1][red]{on line,
arc=7pt,colback=#1!10!white,colframe=#1!50!black,
before upper={\rule[-3pt]{0pt}{10pt}},boxrule=1pt,
boxsep=0pt,left=6pt,right=6pt,top=2pt,bottom=2pt}

\def\arxiv{}

\title{Covert Attacks on Machine Learning Training in Passively Secure MPC}

\ifthenelse{\boolean{public}}{
\author{
  Matthew Jagielski \\
  Google DeepMind \\
  \texttt{jagielski@google.com} \\
  \And
  Daniel Escudero \\
  No Affiliation \\
  \texttt{daniel.escudero@protonmail.com} \\
  \AND
  Rahul Rachuri \\
  Visa Research\thanks{Work done while at Aarhus University.} \\
  \texttt{srachuri@visa.com} \\
  \And
  Peter Scholl \\ 
  Aarhus University \\
  \texttt{peter.scholl@cs.au.dk} \\
}
}
{
\author{Anonymous Authors}
}

\begin{document}

\maketitle

\begin{abstract}
Secure multiparty computation (MPC) allows data owners to train machine learning models on combined data while keeping the underlying training data private. The MPC threat model either considers an adversary who \emph{passively} corrupts some parties without affecting their overall behavior, or an adversary who \emph{actively} modifies the behavior of corrupt parties.
It has been argued that in some settings, active security is not a major concern, partly because of the potential risk of reputation loss if a party is detected cheating.

In this work we show explicit, simple, and effective attacks that an active adversary can run on existing passively secure MPC training protocols, while keeping essentially \emph{zero} risk of the attack being detected. The attacks we show can compromise both the integrity and privacy of the model, including attacks reconstructing exact training data.
Our results challenge the belief that a threat model that does not include malicious behavior by the involved parties may be reasonable in the context of PPML, motivating the use of actively secure protocols for training.
\end{abstract}

\section{Introduction}
\label{sec:introduction}
Secure multiparty computation (MPC) allows data owners to jointly train machine learning (ML) models on pooled data, enabling better models while providing provable privacy protection against colluding servers. 
MPC training protocols can be designed to defend against ``passive'' or ``active'' adversaries. A passive adversary follows the rules of the protocol as expected, and only seeks to learn as much as possible from the protocol communication and the resulting model. An active adversary, by contrast, could deviate arbitrarily from the protocol, motivated perhaps by carefully manipulating the model to misbehave or make it leak other parties' data. Naturally, defending from an active adversary is more ideal but also more challenging than defending from a passive adversary, which manifests itself in protocols that require more computation and communication. Due to these complications, it is very common for research in the area of MPC-based ML to consider passive adversaries for both training and inference~\cite{SP:MohZha17,PoPETS:WagGupCha19,PoPETS:WTBKMR21,EPRINT:KelSun22,EPRINT:LauLipMie06,NDSS:DemSchZoh15,USENIX:ZDCPPS21,EPRINT:CDGGJK21,EPRINT:CGLLR17,SP:PZMZS24,Iron}, and several implemented frameworks only withstand passive adversaries~\cite{ESORICS:BogLauWil08,Syft,TF-Enc,MPyC,Crypten,SP:TKTW21}. 

Although deploying a passive protocol is appealing in practice due to the smaller costs, this requires arguing that the system will not be attacked by an active adversary. One common ``argument'' is that active adversaries can be detected since their attacks lead to large deviations from typical protocol behavior (in machine learning training, this may be done by checking the model's accuracy on test data); if cheating is detected, it would be a reputational risk to the party involved (which is especially significant for established companies, a common setting studied in privacy-preserving ML). Our work challenges this argument, finding that \emph{it is possible for an active adversary to achieve a variety of malicious behaviors in machine learning models without being detected}.
Our findings can be compared to active adversaries in Bitcoin mining, where strategic manipulation of transactions is hard to detect due to network randomness~\cite{SP:DGKLZB20}. 
Similarly, we find parallels with the concrete risks of deploying algorithms with weak privacy guarantees, echoing concerns in differential privacy research~\cite{jagielski2020auditing, nasr2021adversary}.
We expect our work to encourage more careful parameter selection in MPC ML training protocols.


\subsection{Our Contribution}

We challenge common beliefs in favor of passive security in MPC ML training by designing attacks that an active adversary can carry out on several existing passive protocols, \emph{while keeping zero risk}.
Designing these attacks is challenging because, while passive protocols are not intended to be secure under the presence of an active adversary, the specific failures that an active adversary can introduce are difficult to convert into useful attacks---which is partially the reason why passive security is regarded as sufficient.
We overcome this challenge: we show that the somewhat ``limited'' room for attack that passive protocols have is enough to stealthy corrupt the model, compromise fairness, or leak private training data.

Concretely, our work makes the following contributions:
\begin{itemize}[leftmargin=1.5em, itemsep=-2pt, topsep=-1pt]
    \item In Section~\ref{sec:mpc-attacks}, we show novel low level attacks on MPC protocols which allow an adversary to manipulate the secure computation of comparisons and several activation functions. 
    \item In Section~\ref{sec:highlevel}, we bridge the gap between our low level attacks and adversarial ML goals, by constructing three high level attack strategies: gradient shifting, gradient zeroing, and gradient scaling. These strategies attack the full learning algorithm, rather than specific computations, and can be used to easily implement attacks inspired by those in the adversarial ML literature.\footnote{The challenge here lies in showing how to implement these adversarial ML vulnerabilities given very limited room for attack.}
    \item In Section~\ref{sec:advml}, we use our high level attack strategies to instantiate concrete, stealthy data poisoning and training data privacy attacks. In a plaintext simulation, we demonstrate that these attacks are practical on a variety of datasets and models.
\end{itemize}

\section{Background and Related Work}
\label{sec:background}

\subsection{MPC Background}
In MPC, a set of parties have inputs $x_1,\ldots,x_n$, and jointly compute a function $z = f(x_1, \ldots, x_n)$, such that only the output $z$ is revealed.
Most of our work focuses on \emph{outsourced MPC}, where MPC servers train on secret data provided by external clients, preventing the server from tampering with the training data. 
Adversaries in MPC are typically either \emph{passive}, meaning that corrupt parties follow the protocol specification, or \emph{active}, meaning that corrupt parties may deviate arbitrarily.\footnote{An alternative, intermediate notion of \emph{covert} adversary has also been considered~\cite{TCC:AumLin07}, who may try to actively cheat, but is also incentivized to not be detected, due to a risk of being publicly caught.}



\subsection{MPC Computation}
\label{sec:mpc-box}
Many MPC protocols operate by modelling computations via additions and multiplications defined over a finite field $\F_p$ or a ring. However, in order to emulate real-number arithmetic carried out in plaintext machine learning, we use a fixed-point representation of the data. A lot of the operations involved, such as activation functions (\emph{e.g.}~softmax) and divisions, are expensive to perform natively in MPC, leading them to be approximated to ``MPC-friendly" variants which are a lot cheaper to compute.


In our work, we abstract the intricacies of fixed-point arithmetic away, and implement our attacks over floating point values to take advantage of standard ML tooling; we do not believe this will impact our findings, as fixed point approximation in MPC has been shown to not be too lossy relative to floating point arithmetic \cite{SP:KRCGRS20,eprint2024/1953}. Moreover, fixed-point arithmetic may be vulnerable to even stronger attacks due to expanding the attack surface with operations such as truncation. We provide a formal description of our arithmetic model (an ``arithmetic black box'') in Appendix~\ref{app:fpa}.

\textbf{Secret-Sharing.} MPC protocols typically operate over secret-shared values, which cannot be inspected by any party in the computation; we provide a more formal description of secret sharing in the Appendix~\ref{app:fpa}. For our work, secret-sharing prevents the adversary from reading the model parameters and input data, significantly complicating attacks.

\textbf{Adversarial attacks on MPC.} Our goal is to understand how an active adversary can attack passive MPC ML training without detection. However, the power of an active adversary depends on the specific passive protocol. For this reason, we focus on a concrete family of attacks called additive attacks: when two secret-shared values $x, y$ are to be multiplied, an adversary can specify an error $\epsilon$, and the resulting product becomes $x \cdot y + \epsilon$ rather than $x \cdot y$; in other words, the adversary can inject a chosen additive term into the result of a multiplication.\footnote{A key challenge for attacks is that $\epsilon$ must be chosen independently of $x, y$, as they are secret-shared.} Additive attacks apply to nearly all passive protocols and, in fact, some passive protocols are \emph{only} vulnerable to additive attacks~\cite{STOC:GIPST14}. We craft attacks on ML training with only additive attacks, ensuring broad applicability to passive protocols; stronger protocol-specific attacks are an interesting topic for future work.

\subsection{Machine Learning Training in MPC}
\label{sec:logistic-regression}

In our work, we consider a set of machine learning models (logistic regression, SVMs, and neural networks) trained in MPC. Data is assumed to exist in a secret-shared form. 

As in plaintext training, gradient descent is used to optimize model weights. Since gradient descent is a linear function, it can be computed without using a multiplication, in a linear secret-sharing scheme. Therefore, additive attacks are only possible in the computation of the gradient itself.


Our work relies heavily on the specific operations used to compute gradients in MPC. For two-class (i.e. binary) logistic regression, the model parameters $\theta$ consist of a $d$-dimensional vector of weights $w$, a scalar bias $b$, the input $x$ is also a $d$-dimensional vector, and $y\in \lbrace 0, 1\rbrace$ denotes the label. The prediction made by binary logistic regression is a scalar probability, computed as $p(x) = \msf{sigmoid}(w\cdot x + b)$, where $\msf{sigmoid}(z) = 1/(1+e^{-z})$. The loss $\ell$ is known as the cross entropy loss, and we describe below to compute in MPC $\nabla_w\ell(w, b, x, y)$ and $\nabla_b\ell(w, b, x, y)$ for an example $x, y$. Note that all the values are secret-shared, but we omit the notation for clarity.
\begin{wrapfigure}{r}{0.6\textwidth}
  \setlength{\fboxsep}{6pt}
  \setlength{\fboxrule}{1pt}
  \fbox{
  \begin{minipage}{0.55\textwidth} 
    \begin{enumerate}[leftmargin=2em]
        \item Compute the prediction $D=\Pi_{\mathsf{MatMul}}(x, w) + b$.
        \item Compute the probability $P=\Pi_{\mathsf{sigmoid}}(D)$.
        \item Compute the scalar loss derivative $F=P-y$.
        \item Finally, compute $\nabla_w\ell = \Pi_{\mathsf{ElemMul}}(F, x)$ and $\nabla_b\ell = F$.
    \end{enumerate}
    \captionof{figure}{MPC logistic regression gradient} 
    \label{wrap:mpc_calc} 
  \end{minipage}
  }
\end{wrapfigure}

Here, $\Pi_{\msf{MatMul}}$ performs matrix multiplication, $\Pi_{\msf{sigmoid}}$ implements the sigmoid function in MPC, and $\Pi_{\msf{ElemMul}}$ performs elementwise multiplication. This algorithm can be easily extended to $k$-class classification for $k>2$, by replacing $w$ with a $d\times k$ matrix, the biases by a $k$-dimensional vector, and replacing $\Pi_{\msf{sigmoid}}$ with $\Pi_{\msf{softmax}}$, a multiclass extension of sigmoid that we will describe later in more detail. The same basic structure also holds for SVMs as well, instead computing $P=1-\Pi_{\msf{MatMul}}(y, D)$ and $F=\max(0, P)$. Note that SVMs are only standard in binary classification, and by convention use $y\in \lbrace -1, 1\rbrace$. Neural network gradients are computed with backpropagation. For completeness, we describe the MPC implementation of this algorithm in more detail in Appendix~\ref{app:backprop}.

\subsection{Adversarial Machine Learning}
In a \emph{poisoning} attack, an adversary manipulates the model at training time in order to influence the model's predictions. Typically, poisoning attacks are injected by corrupting training data \cite{nelson2008exploiting}, or, in federated learning, by contributing malicious gradients \cite{bagdasaryan2020backdoor}. Poisoning attacks fall into three main attacker objectives. Availability attacks \cite{biggio2012poisoning, jagielski2018manipulating} destroy model accuracy, making them easy to detect and difficult to achieve, requiring many poisoning examples. Backdoor attacks inject a \emph{backdoor trigger} into the model~\cite{gu2017badnets, chen2017targeted}. The poisoned model will classify ``triggered'' out-of-distribution examples consistently with the attack's target class. 
Finally, targeted attacks \cite{USENIX:SMKDD18, shafahi2018poison, jagielski2021subpopulation} change the model's classification on a small set, or a subpopulation of target examples, remaining stealthy by adding a small amount of data and corrupting the model only for a small number of examples.

Membership inference (MI) \cite{SP:SSSS17,CSF:YGFJ18} is a privacy attack where an adversary infers whether a target example was used to train a model. MI attacks generally make a prediction based on the loss on the example, and state of the art attacks \cite{carlini2022membership, ye2022enhanced} \emph{calibrate} this loss to the ``hardness'' of an example. 
MI is a common subroutine in reconstruction attacks~\cite{carlini2021extracting}, which extract entire training examples from a model. High MI success rates are evidence of vulnerability to reconstruction attacks~\cite{balle2022reconstructing, jagielski2023note, kaissis2023bounding}.

\subsection{Related Work}
To the best of our knowledge, we are the first to consider an active adversary in MPC who attacks ML training. Lehmkuhl et al.~\cite{USENIX:LMSP21} consider an active adversary at inference time, showing an active client can use additive attacks to steal a server's model weights. Chaudhari et al.~\cite{EPRINT:ChaJagOpr22} design an MPC protocol to mitigate input-level data poisoning and privacy attacks.

\section{Attacking Activation Functions}
\label{sec:mpc-attacks}

\begin{table}[t] 
\centering
\caption{Overview of manipulations across activation functions. $\Delta$ is a large positive scalar constant, $\delta$ an arbitrary constant, $v$ an arbitrary vector, and $e_i$ the $i$th basis vector of dimension $K$, the number of classes. Lehmkuhl et al.~\cite{USENIX:LMSP21} use input modification on ReLUs for their inference-time attack.}
\label{tab:manipulations}
\begin{tabular}{@{}lccc@{}} 
\toprule
\textbf{Activation Function} & \textbf{Input Modification} & \textbf{Activation Modification} & \textbf{Combined Modifications} \\
\midrule
$\mathbbm{1}(x \ge y)$ & $\lbrace 0, 1\rbrace$ & $1-\mathbbm{1}(x \ge y)$ & $\lbrace 0, 1\rbrace $\\
ReLU & $\lbrace 0, x + \Delta \rbrace$ \cite{USENIX:LMSP21} & $\min(0, x)$ \ & $\lbrace 0, x - \Delta\rbrace$ \\
Piecewise Sigmoid & $\lbrace 0, 1\rbrace$ & Difficult & $\lbrace x + \Delta, x - \Delta\rbrace$\\
Direct Sigmoid & $\lbrace0, 1\rbrace$ & $\text{Sigmoid}(x)+\delta$ & $\lbrace \delta, 1+\delta\rbrace$\\
Direct Softmax & $\lbrace e_i |~i\in [K] \rbrace$ & $\text{Softmax}(x)+v$ & $\lbrace e_i + v~|~ i\in [K]\rbrace$\\
\bottomrule
\end{tabular}
\vspace{-2mm}
\end{table}

In this section, we discuss the implementations of a variety of activation functions in passively secure MPC, and characterize the possible ways that an active adversary can tamper with them.
For each activation function, we show an active adversary can modify the output in three ways: first, by only modifying its input, or the input into the function; second, by keeping the input consistent and only modifying the computation; and third, by modifying both the input and the computation. We remind the reader that these manipulations must be made without knowledge of the honest parties' inputs to the function. In Table~\ref{tab:manipulations}, we offer a high-level overview of the results of the three strategies on the five activation functions we consider. We remark that our overview of different protocols for each activation function is not exhaustive, but we select a popular representative set.

\textbf{Secure Comparison.} The goal of a secure comparison protocol is to obtain a secret-shared bit $b=\mathbbm{1}(x\ge y)$, for secret-shared arithmetic values $x, y$. This is the activation function in SVMs, and is a building block for ReLUs and piecewise linear sigmoids, which we will attack later. Because comparison cannot be represented nicely as an arithmetic circuit, MPC protocols often convert secret-shared integers into Boolean shares, before running a Boolean comparison circuit~\cite{NDSS:DemSchZoh15,C:EGKRS20}.
We observe that, because a comparison circuit has only a single output bit, it is trivial for a malicious adversary to flip the result of a comparison, by injecting an error to the final AND gate of the circuit.

Note that flipping does not allow the adversary to \emph{choose} the comparison result in and of itself. However, if $x$ (or $y$) is the outcome of a multiplication, the adversary can add a large constant offset to $x$, which will result in $x>y$ with high probability, or add a large negative offset to $x$, forcing the output to 0. Combining input modification with result flipping can also force the output to 0 or 1.

\textbf{ReLU.} ReLU can be computed exactly as $\msf{ReLU} (x) = b \cdot x$, where $b$ is computed with a secure comparison $\mathbbm{1}(x\ge 0)$. An adversary who flips the comparison can modify the result to $\min(0, x)=x-\msf{ReLU}(x)$. Lehmkuhl et al.~\cite{USENIX:LMSP21} show how to manipulate ReLU with input modifications if $x$ is the output of a multiplication: adding a large positive constant $\Delta$ can force $x>0$, leading the ReLU to output $x+\Delta$; subtracting $\Delta$ likewise forces the ReLU to output 0. We also can combine input modification with comparison flipping to output either $0$ or $x-\Delta$.

There are two typical ways to compute sigmoid and softmax. An approach that has gained recent popularity due to its simplicity is approximating these complex functions using piecewise polynomials. The advantage of doing so is that these approximations replace exponentials and divisions, which are hard to compute in MPC, with a combination of simpler operations like multiplications and comparisons. However, recent work has also demonstrated the practicality of direct approximations of sigmoid and softmax, by approximating exponentiation. Below, we show attacks on both of these methods.

\textbf{Piecewise Linear Sigmoid} A piecewise linear approximation of sigmoid, used in e.g.~\cite{SP:MohZha17,CCS:MohRin18} can be defined as:

\begin{equation}
	\msf{sigmoid} (x) = \begin{cases}
		 0 & x < - 1/2 \\
		 x + 1/2 & -1/2 \le x \le 1/2 \\
		 1 & x > 1/2
	\end{cases}
\end{equation}

We can compute the approximation via the following equation, $\msf{sigmoid(x)} = (b_1 \oplus 1) \cdot b_2 \cdot (x + 1/2) + (b_2 \oplus 1)$, where $b_1 = \mathbbm{1}(x \le -1/2)$ and $b_2 = \mathbbm{1}(x \le 1/2)$.
Since we operate with secret-shares that are in the 2's complement representation, bits $b_1, b_2$ correspond to the most significant bits of $x + 1/2$ and $x - 1/2$ respectively, and are computed with secure comparison.
Computing the full piecewise polynomial approximation of sigmoid costs two calls to secure comparison, followed by two multiplications of these secret-shared bits with the secret-shared input $x$, over the arithmetic domain.

By tampering with the output of both comparisons, an adversary can force the 0 and $x+1/2$ cases of sigmoid to become 1, and flip the 1 case to become 0, as can be seen in the following case analysis:

\begin{enumerate}
	\item When $x - 1/2 < 0$: The bits $b_1, b_2$ should have both been set to 1 if computed honestly. If the adversary flips both outputs, they will both now become 0 and the sigmoid computation is set to 1, instead of the correct output of 0.
	\item When $-1/2 < x < 1/2$: $b_1$ should have been set to 0, and $b_2$ to 1.
	Flipping $b_1$ and $b_2$ forces the sigmoid output to be 1.
	\item When $x > 1/2$: The bits that should have both been set to 1, will now be set to 0. The sigmoid therefore outputs 0 instead of 1. 
\end{enumerate}

On its own, unless the adversary is able to guess the input $x$, there is little to be gained from only manipulating the piecewise linear sigmoid computation, as there is no universal strategy for producing a given output. However, input manipulations solve this problem. By adding a large positive constant to the input, $x+\Delta>1/2$ will force the output to 1, or to 0 by subtracting: $x-\Delta<-1/2$. By forcing the input into a given region of the piecewise function, manipulating the sigmoid computation can flip either $x+\Delta$ or $x-\Delta$ into the middle range, leading the sigmoid computation to return an out-of-bounds positive or negative value $x+\Delta+1/2$ or $x-\Delta-1/2$ (which we simply write as $x+\Delta$ or $x-\Delta$ in Table~\ref{tab:manipulations}).

\textbf{Direct Computation of Sigmoid and Softmax.} Recall the plaintext computation of softmax and sigmoid:
$$
\text{Softmax}(x)=\frac{\exp(x_i)}{\Sigma_j \exp(x_j)},~~ \text{Sigmoid}(x)=\frac{\exp(x)}{1+\exp(x)},
$$
where $\vec x = (x_1, \ldots, x_l)$ are the linear outputs of the last layer.

Instead of using a piecewise approximation, some work \cite{EPRINT:KelSun22,Crypten,SP:TKTW21} directly approximates both the exponentiation and division in sigmoid and softmax to improve the accuracy of the approximation. Our simplest attack on these computations only exploits the division operation, taking advantage of the computation of $x/y$ as multiplication by the reciprocal of the denominator $x\cdot (1/y)$. Indeed, with only this step, we see that both softmax and sigmoid are the output of a multiplication, and so we can add an arbitrary additive error to their output!

Input modification can in principle result in a wide variety of outputs, but one useful one is to add a large positive constant to a given coordinate $x_i$ (or $x$ itself in sigmoid), so that it dominates the other terms, and results in a vector which is 1 in entry $i$ and 0 otherwise (in sigmoid, this returns 1 or 0). It is also possible to selectively zero out specific entries by adding large negative constants to their input. Combining these input modifications with the computation modification also allows the output to be modified to $e_i + v$ for any basis vector $e_i$ and vector $v$, or $\delta$ or $1+\delta$ for sigmoid. These attacks are those we describe in Table~\ref{tab:manipulations}, but it is also possible to attack the reciprocal and exponentiation in other ways.

The reciprocal is either computed with Newton-Raphson iteration (in \cite{Crypten,SP:TKTW21}) or Goldschmidt's algorithm (in \cite{EPRINT:KelSun22}). Both of these algorithms use a large number of multiplications, which can each be attacked. For example, to compute $1/x$, Newton-Raphson begins with an initial approximation $y_0= 3 \cdot e^{0.5 - x } + 0.003$. Then, the following computation is repeated $y_{n + 1} = y_n \cdot (2 - x \cdot y_n)$, with $n$ set to 10 in practice. As long as $y_0$ is a good initial estimate, $\Lim{n \rightarrow \infty} y_n = \frac{1}{x}$. Any of the multiplications in this protocol can be tampered with to control the output of the reciprocal, and Goldschmidt's algorithm is similarly vulnerable.

It is also possible to attack the exponentiations directly. Some work \cite{Crypten,SP:TKTW21} uses the limit approximation for exponentiation, $\exp(x) = \Lim{n \rightarrow \infty} \left( 1 + \frac{x}{2^n} \right)^{2^n}$. In practice, they use $n = 8$, and use the repeated squaring method to compute the exponentials efficiently. Since this approach involves multiplications, it allows for an adversary to add additive errors to the result.

Keller and Sun \cite{EPRINT:KelSun22} extend the technique of Aly and Smart \cite{EPRINT:AlySma19} to do exponentiation. The high level idea of this technique is to separate the integer and fractional parts of the secret shared value $x$ into $i, r$ such that $\arith{x}^{A} = \arith{i}^{A} + \arith{r}^{B}$, where  $A$ indicates that the integer part is shared over the arithmetic domain and the fractional part, $r$ is shared over the boolean domain ($B$). Then they compute $\arith{2^i}^{A}$ using conventional techniques, and compute $\arith{2^r}^{B}$ using an approximation, such as the Taylor series approximation.

With the following attack strategy, an adversary can make the exponentiation computation output any constant of its choice.
The algorithm from \cite{EPRINT:KelSun22} (Algorithm 2) first computes a bit as,

\begin{equation}
	\ang{z}^B = \Sigma^{k - 1}_{i = 0} 2^{i - f} \ang{x_i}^B < -(k - f - 1)
\end{equation}

where the total bit length of the input $x_i$ is $k$, and the precision is $f$ bits. This bit is used to check if the input $x_i$ is below a certain threshold. Since we use fixed-point representation, instead of computing softmax for very large negative numbers, we set the bit to 1 and output 0 for softmax($x_i$). $\ang{z}^B$ is computed using a comparison operation, and in most cases the database will not have inputs that are huge negative numbers, which means that $z$ will be set to 0 with high probability. Therefore, flipping it sets the bit to 1 in most cases. In Step 13 of Algorithm 2, the final output is computed as $(1 - \msf{Bit2A}(\ang{z}^B)) \cdot \ang{h}^A$, where $\ang{h}^A$ is the output of the algorithm. The idea is that if the input was a large negative value, $(1 - \msf{Bit2A}(\ang{z}^B))$ will be 0, thereby making the final output 0. So by flipping the bit $z$ for one of the inputs $x_i$, with high probability we can make the final output 0. To make the exponentiation output any constant we like, we can make use of the multiplication in Step 13. Since the output is going to be 0, any constant $c$ we add as the additive error at this multiplication will make the final output $c$.

\section{High Level Manipulation Strategies}
\label{sec:highlevel}
Our attacks in Section~\ref{sec:mpc-attacks} manipulate each activation function individually.
Here, we compose these attacks and other additive errors to construct high level manipulations that impact the training algorithm overall.
We treat these manipulations as an API built from the low-level attacks in Section 3, to realize the adversarial goals in Section~\ref{sec:advml}.
Future work targeting other adversarial objectives may directly use these primitives.

A key challenge in designing and using these manipulation strategies is that the adversary does not know the input data or the current model weights, as they are all secret-shared.

\subsection{Gradient Zeroing}
The goal of gradient zeroing is to replace an example's gradient with 0 in as many entries as possible, effectively removing the example from that training step. In logistic regression, we do not know how to achieve this simple goal: the only way to guarantee zeroing would be to force $\msf{sigmoid}(D)=Y$, but this would require a data dependent manipulation, as $Y$ is secret shared. For neural networks, we can force all ReLUs to output 0, which results in nearly all gradient entries being set to 0, except for the final layer biases. In an SVM, we can use this same trick when computing the loss---by adding a large error to $y\cdot D$, we can force the loss, and therefore each entry of the gradient, to 0.

\subsection{Gradient Shifting}
In gradient shifting, an adversary shifts an example gradient $g$ by some fixed vector $v$: $g\rightarrow g+v$. To see why this is possible, recall that each entry of the gradient is computed with a multiplication, such as when computing $\nabla_w\ell$ in the logistic regression gradient in Section~\ref{sec:logistic-regression}. For simplicity, we write the computation of the $i$th entry of the gradient $(\nabla_w\ell)_i$ as $\PMult(F, X_i)$, where $X_i$ is the $i$th feature of the example and $F$ is the example loss. To add a vector $v$ to this gradient computation, the adversary applies an additive error of $v_i$ (the $i$th entry of the shift), changing the computation to $(\nabla_w\ell)_i = \PMult(F, X_i) + v_i$. Note that this is possible only for those gradient entries which are the result of a multiplication, and so only applies to weight gradients, not bias gradients.


The main difficulty that we will run into when using gradient shifting for attacks is that the adversary never knows the model parameters or the data used to compute the clean gradient. Our attacks cannot depend on these values, and the shift $v$ must be effective at targeted a wide variety of possible model parameters. This is especially difficult for neural networks, which have many ``isomorphisms'', wildly different model parameters which correspond to the exact same function \cite{ganju2018property, rolnick2020stealing, jagielski2020high}.

\subsection{Gradient Scaling}
Gradient scaling allows the adversary to scale the gradient by roughly \emph{multiplying} it by a scalar, rather than by shifting by a vector. It is surprising that we can accomplish this at all, because all of our manipulations must be data-independent, and scaling makes a data-dependent change to the gradient. The key property we rely on here is that all gradients are scaled by the derivative of the loss, represented for example by $F$ in the logistic regression gradient algorithm in Section~\ref{sec:logistic-regression}. For logistic regression, while we cannot control the value of $Y$, we do have control over the output of the sigmoid or softmax. By forcing these to arbitrary values, we can scale the gradient.

For logistic regression or neural networks, gradient scaling involves forcing the activation function to a large value. With piecewise linear sigmoid activations, this requires combining input and activation modification. For direct sigmoid or softmax computation, activation computation is sufficient, but may be combined with input modification. Note that we can only force the gradient to encourage strong predictions for a fixed class; we cannot ``amplify'' or ``negate'' a gradient with this approach, as we cannot make the activation's output depend on $Y$.



In SVM training, recall that the loss $F=\max(0, P)$, where $P$ is the margin. Then it is possible to scale gradients by increasing $P$ by a large constant, or by decreasing $P$ and flipping the secure comparison. These will increase the confidence of the SVM on $y=-1$ and $y=1$, respectively.

\section{Attacks on Machine Learning Training}
\label{sec:advml}
We now showcase the variety of adversarial goals which can be achieved with our high level strategies from Section~\ref{sec:highlevel}. Here, we run experiments on logistic regression and neural networks. We use the Fashion MNIST (FMNIST), Census, Purchase, and Texas datasets, as they are standard in the literature. For Census, Purchase, and Texas, we sample 20000 records for the clean training set, allow the adversary 20000 records for their own training dataset, and 20000 records for the test dataset. We train all models with SGD using hyperparameters selected by a grid search on a distinct data split.

\ifx\arxiv\undefined
\begin{wraptable}{r}{0.53\textwidth} %
\else
\begin{table}[] %
\fi
    \centering 
    \caption{A summary of concrete attacks, their capabilities, stealthiness (accuracy-detectable), and data order obliviousness. Privacy attacks that do not know training data order cannot leak arbitrary examples. The primary capability used is listed.}
    \label{tab:all_attacks}
    \begin{tabular}{@{}llcc@{}}
        \toprule
        Attack Goal & Capabilities & Stealthy? & Oblivious? \\
        \midrule
        Shift Poison & Shifting & Yes & Yes \\
        Fairness     & Zeroing  & No  & No \\
        Membership   & Scaling  & Yes & Untargeted \\
        Reconstruct  & Scaling  & No  & Untargeted \\
        \bottomrule
    \end{tabular}
\ifx\arxiv\undefined
\end{wraptable}
\else
\end{table}
\fi

To implement our attacks, we build a plaintext ``simulator'' of MPC over $\mathbb{R}$, using the Jax library \cite{jax2018github}. Our simulator uses appropriate MPC-friendly approximations, and allows the adversary to inject manipulations directly into the appropriate steps of the computation, to ensure faithfulness to how attacks would be carried out in MPC. We use a simulator instead of an MPC implementation as it allows us to 1) run our experiments locally on a GPU (all experiments use a single P100) and 2) take advantage of the autodifferentiation capabilities of Jax. More details can be found in Appendix~\ref{app:simulator}, and we plan to make our implementation public.

We summarize our attacks in Table~\ref{tab:all_attacks}, noting also which capability from Section~\ref{sec:highlevel} they use, whether they degrade model performance, and whether they require knowledge of training data ordering.

\subsection{Parameter Transfer: Poisoning Linear Models with Gradient Shifting}
\label{sec:paramtransfer}
Traditional poisoning attacks use malicious data to exert a specific influence on the model. Our first attack demonstrates how we can replicate this influence with gradient shifting: if a crafted poison example produces a poisoned gradient $g_p$, a gradient shifting attack can introduce this same gradient into the protocol without any data poisoning. This bears some resemblance to poisoning attacks on federated learning (FL) \cite{bagdasaryan2020backdoor, bhagoji2019analyzing}, which show that changing gradients rather than data (known as \emph{model poisoning}) during training leads to very effective attacks.

The key difference between attacks on FL and our attacks is that, in our setting, gradient shifting must be carried out without knowing the current model parameters, as their values are secret shared. Indeed, all work on FL poisoning uses knowledge of the parameters. To circumvent our restriction, we propose a general strategy, \emph{parameter transfer}, to poison models with unknown parameters. In parameter transfer, the adversary trains a ``reference''  model on their own data, computes a poisoned gradient on this reference model, and runs gradient shifting to add this poisoned gradient to the overall gradient.

Neural networks are difficult to attack with parameter transfer due to their nonconvex loss surface: there are many settings of ``good'' weights which are far from each other, including isomorphisms producing identical outputs. Then gradients on one network do not transfer to another, except in fine tuning settings \cite{ilharco2022editing}. We address this shortcoming later, and run parameter transfer on simple models.

\textbf{Evaluation.} We use parameter transfer gradient shifting to implement backdoor and targeted poisoning attacks on softmax-based logistic regression. We train the adversary's reference model and compute transfer gradients with the adversary's disjoint training dataset. The goal of our backdoor attack is, in every dataset, to cause any input from class 0 to be misclassified as class 1 when the first feature is set to a value of 1. The goal of our targeted attacks is to misclassify 5 randomly selected test examples to a randomly selected class. We measure the success rate of backdoor attacks on triggered test examples (distinct from both the clean training and adversary datasets), and the success rate of targeted attacks on the targeted test examples, all averaged over 10 runs.

We report our results for backdoor attacks in Table~\ref{tab:backdoor_param_transfer} and our results for targeted attacks in Table~\ref{tab:target_param_transfer}.
Our attacks achieve near perfect success at each adversarial goal, with little to no performance degradation---model accuracy never drops by more than 2.3\% on average.

\begin{table*}[t]
  \centering
  \caption{Overview of parameter transfer-based gradient shifting attacks, reporting Test accuracy and Attack Success Rate (ASR). Table (a) covers backdoor attacks, while (b) covers targeted attacks.}
  \label{tab:combined_gradient_shifting_attacks}
  
  \begin{subtable}[t]{0.48\textwidth}
    \centering
    \caption{Backdoor attacks via parameter transfer.}
    \label{tab:backdoor_param_transfer}
    \begin{tabular}{@{}ll cc cc@{}}
      \toprule
      & & \multicolumn{2}{c}{No Attack} & \multicolumn{2}{c}{Attack} \\
      \cmidrule(lr){3-4} \cmidrule(lr){5-6}
      Dataset & Model & Test & ASR & Test & ASR \\
      \midrule
      FMNIST   & LR & 0.84 & 0.004 & 0.83 & 0.9996 \\
      Texas    & LR & 0.61 & 0.006 & 0.61 & 1.0000 \\
      Purchase & LR & 0.70 & 0.000 & 0.70 & 1.0000 \\
      \bottomrule
    \end{tabular}
  \end{subtable}
  \hfill 
  \begin{subtable}[t]{0.47\textwidth}
    \centering
    \caption{Targeted attacks via parameter transfer.}
    \label{tab:target_param_transfer}
    \begin{tabular}{@{}ll cc cc@{}}
      \toprule
      & & \multicolumn{2}{c}{No Attack} & \multicolumn{2}{c}{Attack} \\
      \cmidrule(lr){3-4} \cmidrule(lr){5-6}
      Dataset & Model & Test & ASR & Test & ASR \\
      \midrule
      FMNIST   & LR & 0.84 & 0 & 0.83 & 1 \\
      Texas    & LR & 0.60 & 0 & 0.59 & 1 \\
      Purchase & LR & 0.70 & 0 & 0.67 & 1 \\
      \bottomrule
    \end{tabular}
  \end{subtable}
\end{table*}


\begin{table}[t]
    \centering 
    \begin{minipage}[t]{0.45\textwidth}
        \centering 
        \caption{Backdoor attacks with neuron override-based gradient shifting, reporting Test accuracy and Attack Success Rate.}
        \label{tab:backdoor_neuron_override}
        \begin{tabular}{@{}llcccc@{}} 
            \toprule
            \multirow{2}{*}{Dataset} & \multirow{2}{*}{Model} & \multicolumn{2}{c}{No Attack} & \multicolumn{2}{c}{Attack} \\
            \cmidrule(lr){3-4} \cmidrule(lr){5-6} 
            & & Test & ASR & Test & ASR \\
            \midrule
            FMNIST & NN & 0.86 & 0.003 & 0.86 & 0.90 \\
            Texas & NN & 0.58 & 0.01 & 0.57 & 1.00 \\
            Purchase & NN & 0.55 & 0.000 & 0.51 & 1.00 \\
            \bottomrule
        \end{tabular}
    \end{minipage}
    \hfill 
    \begin{minipage}[t]{0.53\textwidth} 
        \centering 
        \caption{True positive rate (TPR) at low false positive rate (FPR) for gradient scaling-based MI amplification. Gradient scaling substantially increases MI risk, sometimes by factors of 20-40x. Confidence intervals ($p<0.01$) for all values have width $<0.26\%$.}
        \label{tab:mi}
        \begin{tabular}{@{}lrrrr@{}}
            \toprule
            \multirow{2}{*}{Data/Model} & \multicolumn{2}{c}{With Scaling} & \multicolumn{2}{c}{No Scaling} \\
            \cmidrule(lr){2-3} \cmidrule(lr){4-5} 
            & TPR@$.1\%$ & $1\%$ & $.1\%$ & $1\%$ \\
            \midrule
            FMNIST/LR     & 4.8\%  & 11.9\% & 0.18\% & 1.5\%  \\
            Texas/LR      & 76.0\% & 93.2\% & 3.0\%  & 10.8\% \\
            Purchase/LR   & 4.5\%  & 22.4\% & 1.9\%  & 11.0\% \\
            \midrule 
            FMNIST/NN     & 2.3\%  & 7.1\%  & 0.2\%  & 1.7\%  \\
            Texas/NN      & 46.6\% & 75.3\% & 0.8\%  & 4.3\%  \\
            Purchase/NN   & 6.0\%  & 13.6\% & 0.4\%  & 2.9\%  \\
            \bottomrule
        \end{tabular}
    \end{minipage}
\end{table}

\subsection{Neuron Override: Poisoning Neural Networks with Gradient Shifting}
\label{sec:neuronoverride}
Neural networks are not vulnerable to parameter transfer due to their nonconvex loss landscapes. We instead design an attack called \emph{neuron override}. In neuron override, the adversary concentrates the backdoor attack into a single neuron, forcing it to activate only on backdoored data, and connects this neuron's activation to the output corresponding to the target class. We implement neuron override specifically for backdoor attacks, but believe it can also be used for targeted attacks.

Concretely, for a two layer ReLU neural network, without loss of generality, we override the first neuron. We use gradient shifting to add a shift of $\delta - c\mu$ to the first neuron's input weights. Here, $\delta$ is the backdoor trigger direction, a direction we expect only backdoored examples to be highly correlated with. We set $\mu$ as the average of some nontargeted data, which we use to reduce the number of standard examples that activate the neuron, and $c$ is a hyperparameter used to trade off these contributions. In the second layer, we also shift the weight from this neuron to the target class by a large positive value, to force examples activating this neuron to be classified into the target class. This strategy is inspired by the handcrafted backdoors of Hong et al.~\cite{hong2022handcrafted}, although their attacks had the advantage of being able to inspect model weights when introducing the backdoor; our attacks must override a neuron regardless of its original weights.

\paragraph{Evaluation:}
We evaluate our neuron override attacks on the same backdoor tasks we considered for parameter transfer. We attack softmax-based two layer neural network models with varying hidden layer sizes, and again measure backdoor attack success rate on triggered test examples, and average over 10 runs. We report our results in Table~\ref{tab:backdoor_neuron_override}. Our attacks lead to limited test accuracy decrease, of no more than 4\% accuracy, but our attacks lead to very high attack success rates.

\subsection{Amplifying Membership Inference with Gradient Scaling}
\label{sec:mi}
Our prior attacks compromise the \emph{integrity} of learned models. We now turn our attention to privacy attacks, directly compromising MPC's privacy guarantees only by deviating from the protocol.

We begin with membership inference (MI). We take advantage of a property identified by recent work, that it is easier to infer membership of ``outlier'' examples, such as mislabeled or noisy examples, compared to more natural ``inlier'' examples~\cite{carlini2022membership}. Our attack uses gradient scaling to ``convert'' any example into an outlier example. Recall that gradient scaling can modify the gradient to encourage an example be predicted as any target class. By selecting any example, and any random target class, our attack uses gradient scaling to force that example to be predicted as the target class. Because it is unlikely that the example will naturally belong to that class, the update is made as if the target were a mislabeled ``outlier'', leaving it vulnerable to MI!

\textbf{Evaluation.} We run our attack on softmax-based logistic regression and neural network models. We run a small amount of tuning to select gradient scaling strengths that result in low accuracy drop, and use a target class of 0 for all gradient scaling.
To run the attack, we adapt the state-of-the-art offline LiRA attack \cite{carlini2022membership}. In LiRA, the adversary trains ``shadow models'' without scaling on their own data. Then, when running the attack on a given model and example, the adversary measures the deviation of the model's predictions on the target example from the predictions of the shadow models. Traditionally, LiRA measures the deviations in the probabilities of the example's correct class, but we account for our scaling by instead measuring the deviation for the gradient scaling target class.

We measure the effectiveness of our attacks in Table~\ref{tab:mi} using True Positive Rate (TPR) at low False Positive Rates (FPRs), a standard measure of MI success, reflecting the attack's precision. We find that our gradient scaling strategy is very effective, improving TPR by a factor of at least 2 and sometimes as large as 40! At these levels, the adversary can be nearly certain that the identified examples are members. Our attacks often have limited impact on model accuracy, such as a 0-2\% decrease on FMNIST models, but can be substantial, such as an 8\% drop on average for NN on Purchase. Attack success and accuracy can be traded off by modifying the gradient scaling strength.

\subsection{Training Data Reconstruction with Gradient Scaling}
\label{sec:recon}
Having established strong membership inference leakage, we now turn to using gradient scaling to perform training data reconstruction. Our reconstruction attack relies on a property exploited by Boenisch et al.~\cite{boenisch2023curious} in the context of federated learning attacks --- the gradient of the first layer of a neural network (or just the weights in a linear model) is a scalar multiple of the input features. This can be seen, for example, in the computation of the gradient $G$ in the protocol for logistic regression in Section~\ref{sec:logistic-regression}; to compute the gradient, the scalar loss derivative $F$ is multiplied by the input $X$.

In MPC, the adversary cannot access the gradients directly, as is possible in (vanilla) federated learning. To still be able to perform reconstruction, we will instead ``overwrite'' the model weights with a highly scaled gradient. For example, using a large scaling of 10000 for a single example in the final batch will result in the original model weights being small in magnitude relative to the scaled gradient; the first layer weights will be a vector which can be rescaled to obtain the target example. Of course, such a large gradient will destroy the original model weights, resulting in an unusable model and an obvious attack, but an adversary may consider running this attack if the possibility of reputation damage is outweighed by the learning sensitive information.

\textbf{Evaluation.}
We run our attack on softmax-based logistic regression and neural networks for Fashion MNIST, using a scaling factor of 10000 for class 0. We attack a random example in the final batch of training for simplicity, but expect that the attack could be carried out in different training steps if desired. After performing the attack, we must select the weights to rescale to recover the reconstructed image. For logistic regression, we use the model weights leading to the class 0 output, and for the neural network, we inspect a small number of neuron weights to find one which contains a coherent image, and rescale these weights to reconstruct the example.

\begin{figure}
    \centering
    \includegraphics[width=0.7\linewidth]{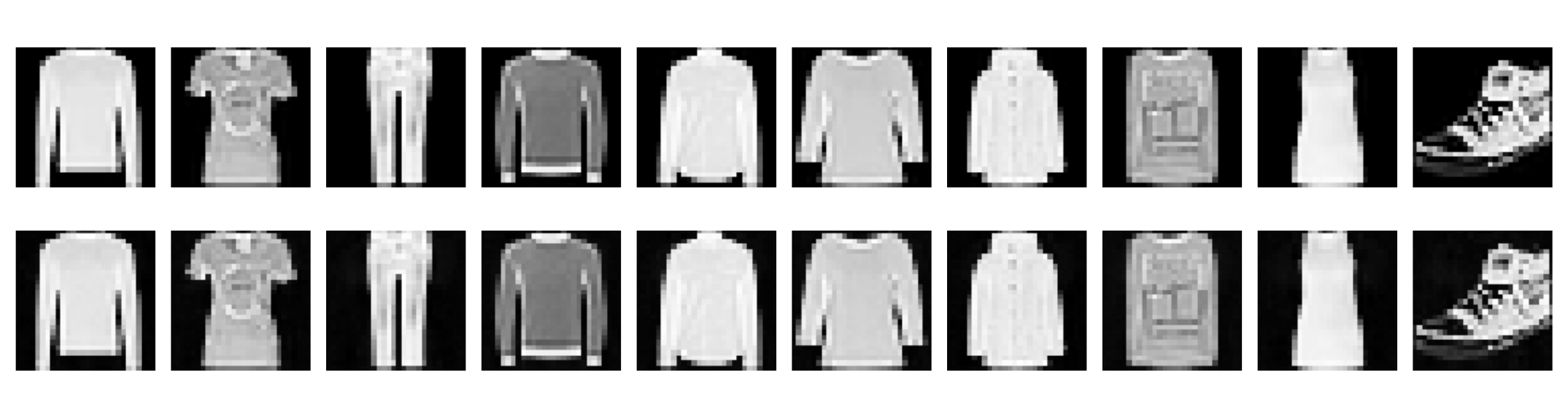}
    \caption{Gradient scaling leads to successful reconstruction attacks on logistic regression. The top row is the original image, and the bottom is its reconstruction.}
    \label{fig:recon_lr_10k}
\end{figure}

\begin{figure}
    \centering
    \includegraphics[width=\linewidth]{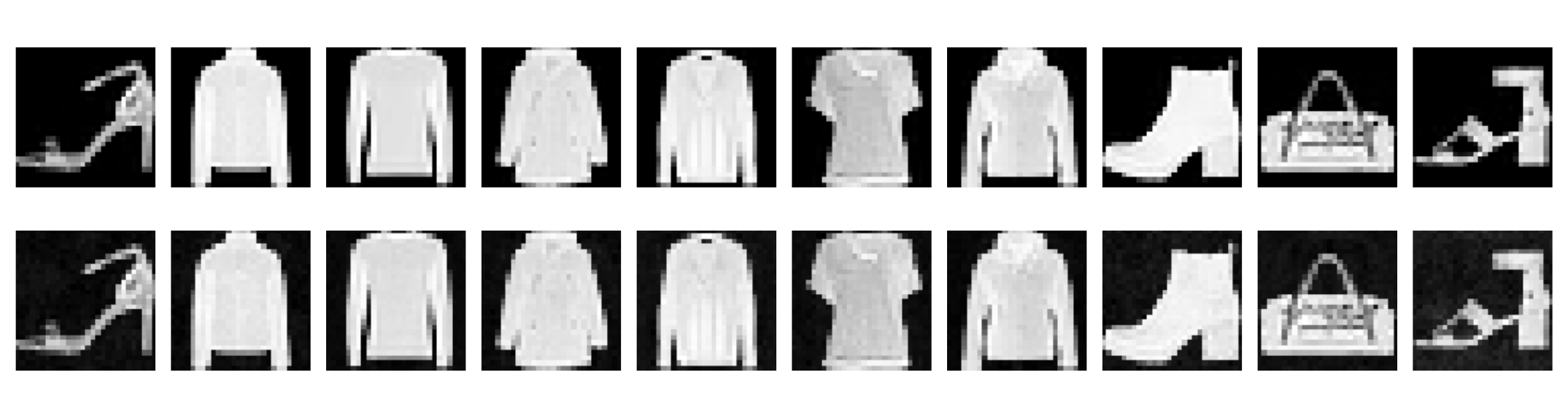}
    \caption{Gradient scaling attacks permit successful reconstruction attacks on neural networks. The top row is the original image, and the bottom row is the reconstructed image.}
    \label{fig:recon_nn}
\end{figure}

We run 10 trials of our attack and present our reconstructions in Figure~\ref{fig:recon_lr_10k} and Figure~\ref{fig:recon_nn} for logistic regression and neural networks, respectively. Visually, they are nearly identical to the original images. The mean absolute error between the original and reconstructed images is 0.017 for logistic regression and 0.036 for neural networks. That is, the average pixel of a reconstruction differs from the original by only 1.7\% (or 3.6\%) of the total pixel range of [0, 255]. However, we reiterate that this attack is not stealthy, reducing model accuracy to 10\% or ``random guessing''.

\subsection{Reducing Fairness with Gradient Zeroing or Scaling}
\label{app:fairness}
In high stakes applications, it is important to build models which do not discriminate on societally relevant attributes, such as race or gender. Multiparty machine learning may even be a way of reducing such unfairness; often, in multiparty settings, each party's data has a different distribution, and, by pooling data, the model improves on all represented subpopulations. We consider such a setting, and show how an adversary can attack the protocol to cause the model's accuracy to degrade for one contributing party, by nullifying the contributions of that party's data to the model.

The most natural way to achieve this aim is to prevent the target party's data from \emph{ever} contributing to the model with gradient zeroing. This requires the adversary to know which examples came from the target party, and so requires knowledge of the data order. Gradient zeroing is not possible for logistic regression, so we also experiment with gradient scaling, to encourage all examples from the target party to be misclassified to a fixed target class. These attacks can be seen as an adaptation to our threat model of attacks considering the implications of removing data \cite{zhang2023forgotten} or poisoning attacks \cite{solans2020poisoning, chang2020adversarial, jagielski2021subpopulation} on model fairness.

\textbf{Evaluation.}
We run this attack using a simulated multiparty setting on the US Census dataset. Using the folktables Python package, we generate five parties by selecting 1000 examples from each of five different, arbitrarily chosen states: CA, OK, OR, WA, and WI. The parties collaborate on a sigmoid-based logistic regression model or a sigmoid-based neural network with 50 neurons in the hidden layer. We train both for 10 epochs with learning rate 0.001 and batch size 100. The adversary targets CA, and performs gradient zeroing or gradient scaling with strength 2 towards class 1; we average all results over 10 trials. We choose this setup as a real-world example of non-iid data, although our results should only improve in more highly non-iid settings.

We report the results of our experiment in Table~\ref{tab:fairness_table}. We report test accuracy for the target state and the average for the other states', both with and without each attack. We find that our attacks all have a disparate impact on model accuracy: the decrease in accuracy on the target state is higher than the decrease in accuracy on the other states. For example, attacking the neural network with gradient scaling decreases accuracy in CA by 7.2\%, while the other states are only harmed by 2.7\%.

\begin{table}[]
    \centering
    \begin{tabular}{c|cccc}
        \multirow{2}{*}{Model/Attack} & \multicolumn{2}{c|}{No Attack} & \multicolumn{2}{c}{Attack} \\
        ~ &  Target & Other & Target & Other \\
        \hline
        LR + Boost & 0.748 & 0.753 & 0.673 & 0.731 \\
        NN + Boost & 0.724 & 0.737 & 0.652 & 0.710 \\
        NN + Zeroing & 0.75 & 0.76 & 0.714 & 0.731 \\
    \end{tabular}
    \caption{Attacks on fairness. Our attacks reduce attack performance disproportionately on the target subpopulation, harming the fairness of the trained model, by preventing one party's contributions to the model.}
    \label{tab:fairness_table}
\end{table}

\subsection{Amplifying Data Poisoning with Gradient Scaling}
\label{app:poisoning}

Parameter transfer and neuron override both use gradient shifting to induce successful poisoning. To add to this, we also show that an adversary can amplify data poisoning attacks by colluding with the malicious data contributors, by scaling the gradients of these malicious examples. This can lead to poisoning that is more effective than the data poisoning alone, allowing attacks to use smaller amounts of poisoning data.

To instantiate this data poisoning amplification, note that the adversary's collusion with malicious data means they know the $Y$ value for the poisoning data, and gradient scaling can be performed to amplify the gradient towards this class. For example, in softmax-based models, the adversary can add a large negative value to the softmax output for the poisoning $Y$. In sigmoid-based models, if $Y=1$, the sigmoid can be tampered with to output a large value, or a small value if $Y=0$.

\textbf{Evaluation.}
We report results for targeted data poisoning here, as we find that amplifying backdoor data poisoning leads to inaccurate models, likely due to the relatively large amount of poisoning data we need to encourage generalization of the backdoor trigger. We attack softmax-based models, including both logistic regression and neural networks. For targeted data, we again target 5 examples, and add between 2 and 10 flipped label training examples per target, and gradient scale by shifting the value of softmax for the target class by between -2 and -5. These hyperparameters depend on the dataset and model, and we perform a small amount of tuning to find these values, identifying a number of training examples which can be added per target which lead to a weak attack that can be improved with scaling. If we used more data per target, the attacks would not need scaling. In practice, the amount of data per target is more likely to be a constraint of the adversary, which would be set independently, and offline experimentation could be used to set the scaling strength.

We report our results in Table~\ref{tab:target_grad_scaling}. As expected, gradient scaling allows our attacks to be more effective at the data poisoning levels we test. Of course, stronger data poisoning would allow weaker scaling to be effective, but our attacks can enable strong attacks when they were impossible at small data poisoning strengths. Our attacks also generally have small impact on model performance, although neural networks on Fashion MNIST and Purchase see substantial accuracy decreases. In practice, an adversary can test hyperparameter values offline in order to understand how aggressive they can afford to be given a fixed accuracy constraint.

\begin{table}[]
    \centering
    \begin{tabular}{c|ccc}
        Dataset/Model & No Attack & No Scale & Scale \\
        \hline
        FMNIST/LR & 0.84/0.02 & 0.84/0.04 & 0.83/0.72 \\
        Texas/LR & 0.61/0 & 0.60/0.08 & 0.60/1 \\
        Purchase/LR & 0.70/0 & 0.70/0 & 0.69/0.72 \\
        FMNIST/NN & 0.86/0 & 0.86/0.36 & 0.75/0.52 \\
        Texas/NN & 0.59/0 & 0.58/0.28 & 0.56/0.94 \\
        Purchase/NN & 0.55/0.02 & 0.54/0.36 & 0.41/0.9\\
        \end{tabular}
    \caption{Test accuracy/attack success rate for targeted attacks with gradient scaling. Gradient scaling results in a stronger attack than could be achieved without scaling.}
    \label{tab:target_grad_scaling}
\end{table}

\subsection{Other Possible Attacks and Possible Mitigations}
We have instantiated each adversarial goal with one main high level strategy, but some goals may be achievable with multiple strategies.

\textbf{Availability Poisoning.} These attacks are possible with nearly any large, untargeted, modification to the model. Hypothetical approaches are heavy gradient scaling of many example gradients, batch-wise gradient zeroing during training, to ensure that the model parameters never change during training, or gradient shifting with large modifications, which we briefly evaluate in Table~\ref{tab:avail_param_transfer}.

\begin{table}[]
    \centering
    \begin{tabular}{cc|cc}
        Dataset & Model & No Attack & Attack \\
        \hline
        FMNIST & LR & .836 & .717 \\
        Texas & LR & .605 & .565 \\
    \end{tabular}
    \caption{Availability attacks with parameter transfer-based gradient shifting. We use a small shift gradient, so our performance degradation can be improved significantly by changing this value.}
    \label{tab:avail_param_transfer}
\end{table}

\textbf{Membership Inference.} It has been shown that MI can be amplified through the inclusion or exclusion of other examples \cite{chen2022amplifying, tramer2022truth, carlini2022privacy}. An adversary could replicate these indirect attacks with any of gradient zeroing, shifting, or scaling.

\section{Potential Protocol-Level Mitigations}
\label{sec:mitigations}

An obvious mitigation, immediately preventing all of our attacks, is to use actively secure MPC protocols for training. Of course, such protocols come with a significant performance (in terms of run time) hit. This motivates the question of whether one could design “middle ground” mitigations, which can prevent attack, such as the ones proposed in our work, without sacrificing too much performance relative to the passive protocol. Towards this, we outline possible design strategies to mitigate these attacks. We would like to emphasise that adaptive attacks may be able to circumvent them, and more careful analysis is required.

Gradient scaling requires manipulating the loss computation, and can be mitigated by computing the loss given the logits with an actively secure sigmoid or softmax. Computing these activation functions takes a small fraction of training time, as they are performed once per example, which does not scale with the number of parameters in the model. In \cite{EPRINT:KelSun22}, softmax computation takes less than one percent of compute time, making this a cheap but powerful mitigation. This approach would fully stop our boosting of poisoning attacks, attacks compromising fairness, and membership inference attacks. 

Mitigating gradient shifting attacks is more challenging, as multiplications in backpropagation can be independently tampered with. However, the gradient shifting attacks we run all requires tampering with many multiplications, so one potential avenue to reduce the risk of an attack is to somehow check only an $\alpha$ fraction of multiplications, rather than all of them. We do not know how to design such a protocol, as naive approaches would either not improve efficiency, or invalidate the goal by allowing the adversary to know in advance which multiplications will be checked. However, if possible, and the adversary introduces $p$ additive errors throughout the protocol, then the probability that the adversary is not caught is $(1-\alpha)^p$. Our gradient shifting poisoning attacks on FMNIST, for example, modify roughly 1.5 million multiplications in total, which is large enough that even checking an $\alpha < 2 \cdot 10^{-5}$ fraction of multiplications will still catch the adversary with $> 1 - 2^{40}$ probability. However, our attacks are not optimized to reduce the number of additive errors, and it is likely these attacks can remain effective with orders of magnitude fewer errors by, for example, making larger, less frequent, or more sparse modifications.
\section{Discussion}
Our results show that malicious adversaries can stealthily corrupt models trained with passive MPC protocols. This challenges the perspective that attacks on such protocols will be easy to prevent with reputation- or incentive-based approaches. 
We hope that our work promotes more research at the intersection of MPC and adversarial ML, in addition to having takeaways for each:

\textbf{For the MPC community.} We show that in ML training, if an active adversary is given access to the output of the computation, privacy can be completely broken in the passive setting. An interesting future theoretical question is to understand if there exists a broader class of functionalities with such privacy issues.
This is analogous to related-key attacks on AES \cite{AC:BelCasMil11,C:ABPP14}, where adversaries exploit structural relationships between keys to induce predictable variations in internal states, ultimately enabling key recovery. In both cases, the attacker leverages correlated inputs to bypass standard security assumptions.
We also hope to encourage exploration into ``middle ground" protocols between the two models, which defend against well defined classes of attacks, as we discuss in Section~\ref{sec:mitigations}. Our work underscores the importance of actively secure ML training; we hope it provides useful information for MPC practitioners when evaluating threat models for applications. 

\textbf{For the adversarial ML community.} The ``additive error'' threat model we consider in our work may be interesting to investigate more deeply in adversarial ML. It may be possible to improve the effectiveness and stealthiness of our attacks or understand the threat model from a theoretical perspective. Beyond this threat model, we encourage adversarial ML research to consider MPC as a setting to propose attacks, consider new threat models, and design concretely efficient defenses.

\textbf{Limitations.} One limitation of our work is that we work over floating point arithmetic, rather than fixed-point; we do not believe this significantly impacts findings \cite{SP:KRCGRS20,eprint2024/1953}. Our experiments are limited to small models; due to the cost of training in MPC, the literature tends to use very small models \cite{EPRINT:KelSun22}, and we expect our attacks to generalize to larger models.


\subsection*{Acknowledgements}
The work of P. Scholl was supported in part by grants from the Digital Research Fund Denmark (DIREC), the Danish
Independent Research Council under Grant-ID DFF-0165-00107B (C3PO) and by Cyberagentur under the project Encrypted Computing.

\bibliographystyle{IEEEtran}
\bibliography{references,abbrev3,shrunk,advmlref}

\appendix
\section{Instantiating the Attacks in MPC}

\subsection{MPC Over Fixed-Point Arithmetic}
\label{app:fpa}

An MPC protocol is often modelled abstractly as a secure ``arithmetic black box'' (ABB) --- that is, a box that receives and stores private inputs from clients, and when instructed by all parties, performs computations on the inputs and reveals any desired outputs.
While classic MPC protocols model computations via additions and multiplications over some finite field or ring, in machine learning we aim to emulate real-number arithmetic, so instead assume a fixed-point representation of the data.
In fixed-point arithmetic with precision $f \in \mathbb{N}$, real numbers are approximated as rationals of the form $x/2^f$, where $x \in \Z$.
When emulating fixed-point arithmetic in an MPC protocol, we require $x$ to lie in the ring $\Z_M$ for some modulus $M$, and assume that $M$ is large enough such that during each addition or multiplication in the computation, there is no wraparound modulo $M$ so we avoid overflow.
The MPC protocol then emulates fixed-point arithmetic via additions and multiplications in $\Z_M$, together with a specialized protocol to truncate the result after each fixed-point multiplication.
In Figure~\ref{fig:mpc-box}, we describe the arithmetic black box functionality, that captures fixed-point arithmetic while allowing an adversary to introduce an additive error after each multiplication operation.
Note that the functionality assumes that any fixed-point value $x/2^f$ input by the parties in the $\mathsf{Input}, \mathsf{Store}$ or $\mathsf{LinComb}$ steps is represented solely by the value $x \in \Z_M$.

\begin{protofig}{$\FMPC$}{MPC arithmetic black box for fixed-point arithmetic modulo $M$}{mpc-box}

  The functionality interacts with a set of clients $\mathcal{C} = \{C_1,\ldots,C_m\}$ and a set of parties $\mathcal{P} = \{P_1,\ldots,P_n\}$, and is parameterized by a precision value $f$.

  \begin{itemize}
  \item
    On receiving $(\mathsf{Input},\mathtt{id},C_i)$ from all parties and clients, and $(\mathsf{Input},\mathtt{id},C_i, x)$ from $C_i$, where $x\in \Z_M$, store $(\mathtt{id}, x)$.
  \item
    On receiving $(\mathsf{Store},\mathtt{id},c)$ from all parties, where $c\in \Z_M$, store $(\mathtt{id}, c)$.
  \item
    On receiving $(\mathsf{LinComb}, c_1, c_2, \mathtt{id}_{1},\mathtt{id}_2, \mathtt{id}_3)$ from all parties, where $c_1, c_2 \in \Z_M$, retrieve $(\mathtt{id}_1, x)$, $(\mathtt{id}_2, y)$ and store $(\mathtt{id}_3, z)$, where $z = \lfloor (c_1 \cdot x + c_2 \cdot y)/2^f \rfloor$. 
  \item
    On receiving $(\mathsf{Mult},\mathtt{id}_1, \mathtt{id}_{2}, \mathtt{id}_3)$ from all parties, and receiving $(\mathsf{Error},\epsilon)$ from the adversary, where $\epsilon \in \Z_M$, retrieve $(\mathtt{id}_1, x)$, $(\mathtt{id}_2, y)$ and store $(\mathtt{id}_3, z)$, where $z = \lfloor x \cdot y/2^f \rfloor + \epsilon$.
  \item
    On receiving $(\mathsf{Open},\mathtt{id}, P_i)$ from all parties, retrieve $(\mathtt{id}, x)$ and send it to party $P_i$.
  \end{itemize}
\end{protofig}

\paragraph{Secret-Sharing.} Our attacks apply to most MPC protocols based on a linear secret-sharing scheme.
In linear secret sharing, a secret $x \in \Z_M$ is divided into $n$ shares, $x_1, \dots, x_n$, distributed among $n$ parties such that each party $P_i$ holds the share $x_i$.
We use the notation $\sh{x}$ to denote that $x$ is secret-shared across the parties in an MPC protocol.
A secret can be reconstructed, given all $n$ shares, in a linear manner via $x = \sum_i \lambda_i x_i$, where each $\lambda_i \in \Z_M$ is a public reconstruction coefficient.
In the simplest example of additive sharing, each $\lambda_i = 1$ and all $n$ shares are required to reconstruct.
Another example is Shamir secret sharing, where the $i$-th share is an evaluation $p(i)$ of a polynomial $p(X)$ such that $p(0) = x$, and the $\lambda_i$'s are Lagrange coefficients used to interpolate $x$.
Here, the number shares needed for reconstruction is $t+1$, where $t$ is the degree of $p(X)$.

In these schemes, additions are ``free'' because parties can locally add their shares of $\sh{x}, \sh{y}$ to get $\sh{x + y}$.
Multiplications, however, require interaction, and reducing the communication needed to perform secret-shared multiplications is a popular area of research, and especially important for settings like PPML where there are a large number of multiplications to compute.

Our attacks are focused on honest majority protocols using \emph{robust} secret sharing schemes such as Shamir sharing or replicated secret sharing, where the reconstruction process allows parties to easily verify that the correct secret was recovered.
These schemes are often used in practical honest-majority protocols (even in the semi-honest model).
However, we highlight that all of our attacks can also be applied to MPC protocols based on any linear secret-sharing scheme, including additive secret sharing, which is usually used in the dishonest majority setting.










\subsection{Backpropagation}
\label{app:backprop}
Neural networks for classification typically use sigmoid or softmax outputs as in logistic regression, making their gradient computations similar to logistic regression. However, there is added complexity due to their depth. Here we describe how a two layer neural network is computed, as extending this further follows the same principles. First, a two layer neural network has two weight matrices, the $m \times d$ input matrix $w_0$ and the $k\times m$ output matrix $w_1$, and two bias vectors, the $m$ dimension $b_0$ and the $k$ dimension $b_1$, where again $k$ is the number of classes (or 1 in binary classification), and here $m$ is the dimension of the ``hidden layer''. The network prediction $D$ is instead computed as $D=w_1 \text{ReLU}(w_0\cdot x + b_0) + b_1$, where ReLU is a nonlinear activation function $\text{ReLU}(z)=\max(0, z)$. We can write the ReLU inputs $D_0=w_0\cdot x + b_0$, and the outputs from ReLU, $A=\text{ReLU}(D_0)$, are known as the ``activations'' of the first layer, as each entry, or ``neuron'' in this layer is ``activated'' when the ReLU is nonzero. The gradient computation is also slightly more complicated, relying on the backpropagation algorithm~\cite{rumelhart1986learning}. The loss derivative is identical, and the last layer gradients are analogous to logistic regression: $\nabla_{w_1}\ell = \Pi_{\msf{ElemMul}}(F, A)$ and $\nabla_{b_1}\ell = F$. Backpropagation computes the input layer gradients with the chain rule: $\nabla_{b_0}\ell=\Pi_{\msf{ElemMul}}(\Pi_{\msf{MatMul}}(w_1^T, F), \text{dReLU}(D_0))$ and $\nabla_{w_0}\ell = \Pi_{\msf{MatMul}}(\nabla_{b_0}\ell, x)$, where dReLU is the derivative of ReLU, $\text{dReLU}(z)=\mathbbm{1}(z>0)$.

\subsection{Algorithms for Our Attacks}
We present algorithmic descriptions of our attacks, with the adversary's modifications to the protocol highlighted in {\color{blue}{blue}}. Each attack has some preprocessing or postprocessing phase, where the adversary operates entirely in plaintext computation. During the protocol, all values are secret shared, except for the adversary's modifications, which modify the secret shared values.

Our algorithm for Parameter Transfer is in Figure~\ref{fig:paramtransfer} for binary logistic regression. Neuron override attacks are described in Figure~\ref{fig:neuronoverride}, but omitting the majority of the gradient computation for brevity. Privacy attacks are described in Figure~\ref{fig:gradscaleprivacy} - the underlying gradient scaling mechanism is the same, but the adversary's postprocessing (and input parameters) vary between the two attacks.

We also open source our code to improve reproducibility.

\begin{protofig}{Parameter Transfer}{Parameter Transfer Attack from Section~\ref{sec:paramtransfer}}{paramtransfer}

\textbf{Input:} Attacker's local weights $\theta_P$, attacker goal data $X_p, Y_p$. Secret shared batch $X_B, Y_B$, secret shared model weights $\theta_i=(w_i, b_i)$

\textbf{Attacker Preprocessing:}
\begin{enumerate}
    \item Locally compute $G_w = \nabla_w(\theta_P, X_P, Y_P)$.
\end{enumerate}

\textbf{Protocol:}
\begin{enumerate}
    \item Compute prediction $D=\Pi_{\msf{MatMul}}(X_B, w_i) + b_i$.
    \item Compute probability $P=\Pi_{\msf{sigmoid}}(D)$.
    \item Compute scalar loss derivative $F=P-Y_B$.
    \item Finally, compute $\nabla_w\ell = \Pi_{\msf{ElemMul}}(F, X_B) \color{blue}{+ G_w}$ and $\nabla_b\ell = F$.
\end{enumerate}

\end{protofig}

\begin{protofig}{Neuron Override}{Neuron Override Attack from Section~\ref{sec:neuronoverride}}{neuronoverride}

\textbf{Input:} Trigger $\delta$, data mean $\mu$, mean scaling $\alpha$, target neuron index $j$. Secret shared batch $X_B, Y_B$, secret shared model weights $\theta_i=(W^0_i, B^0_i, W^1_i, B^1_i)$

\textbf{Attacker Preprocessing:}
Attacker locally computes
\[
G_{W^0}[i] = \begin{cases} 
          \delta - \alpha\mu & i=j~(\text{target neuron}) \\
          0 & i\neq j~(\text{not target neuron})
       \end{cases}
       \]
       
\[
G_{W^1}[i] = \begin{cases} 
          1 & i=j~(\text{target neuron}) \\
          0 & i\neq j~(\text{not target neuron})
       \end{cases}
       \]

\textbf{Protocol:}
\begin{enumerate}
    \item Following standard backpropagation, adversary shifts hidden layer derivative $\nabla_{W^1}\ell - \color{blue}{G_{W^1}}$ and first layer derivative $\nabla_{W^0}\ell - \color{blue}{G_{W^0}}$.
\end{enumerate}

\end{protofig}

\begin{protofig}{Scaling for Reconstruction Attacks}{Gradient Scaling Attack from Section~\ref{sec:recon}}{gradscaleprivacy}

\textbf{Input:} Attacker protocol inputs: Attacker target class (one hot vector) $y$, scaling strength $C$. Secret shared protocol inputs: Secret shared batch $X_B, Y_B$, secret shared model weights $\theta_i=(w_i, b_i)$.

\textbf{Protocol (last step of training):}
\begin{enumerate}
    \item Compute prediction $D=\Pi_{\msf{MatMul}}(X_B, w_i) + b_i$.
    \item Compute probability $P=\Pi_{\msf{softmax}}(D) \color{blue}{+ C \cdot y}$.
    \item Compute scalar loss derivative $F={\color{blue}{P}}-Y_B$.
    \item Finally, compute $\nabla_w\ell = \Pi_{\msf{ElemMul}}({\color{blue}{F}}, X_B)$ and $\nabla_b\ell = \color{blue}{F}$.
\end{enumerate}

\textbf{After Protocol (Reconstruction):}
\begin{enumerate}
    \item Attack returns $w_y$, a reconstructed input.
\end{enumerate}


\end{protofig}

\subsection{Simulating MPC over reals}
\label{app:simulator}
All of our experiments are done in plaintext, using a framework we build using Jax \cite{jax2018github} to simulate the errors we use in our work as faithfully as possible while computing in floating point arithmetic. For example, we implement the MPC-friendly sigmoid as is proposed in SecureML, with comparison operations, to ensure we can properly attack these operations. We then use Jax's \codeword{custom_vjp} function to replace the gradient of this value with the gradient as computed in SecureML. This allows us to be faithful to the MPC-friendly approximations while simultaneously using Jax to autodifferentiate and accelerate our code on GPU. 

In the implementation of each function, we also add extra parameters for the adversary's modifications. For example, our piecewise linear sigmoid implementation is:

\begin{verbatim}
def mpc_sigmoid(inp, flip_b0=0, flip_b1=0):
    ge_minus_p5 = (inp > -.5) ^ flip_b0
    le_plus_p5 = (inp > .5) ^ flip_b1
    comp1 = jnp.where(ge_minus_p5, inp+.5, 0)
    out = jnp.where(le_plus_p5, 1, comp1)
    return out
\end{verbatim}

Here, \codeword{inp} is an array, representing the batch of inputs to the sigmoid, and the ``flip'' parameters will flip the results of comparisons made in the MPC sigmoid. To flip only one element of a batch, the corresponding ``flip'' parameter should be another array with the same size as \codeword{inp}. Alternatively, setting \codeword{flip_b0} to 0 allows the Jax to ``broadcast'' the computation, applying the same value to each element of the batch (in this case, by not flipping any of the comparisons made in the batch).

To perform a gradient scaling attack, for example, we must modify the inputs to a sigmoid as well. To do this, rather than modifying the code of \codeword{mpc_sigmoid}, we instead modify the prior operation. This allows us to most easily take advantage of Jax's automatic differentiation capabilities with our custom operations, chaining together our ``attackable'' operations to build an end-to-end differentiable and ``attackable'' model. We plan to open source our implementation to accommodate future research.

\end{document}